\documentclass[pdflatex,sn-nature]{sn-jnl}
\usepackage{type1cm}
\usepackage{mathrsfs}
\usepackage{makecell}
\usepackage{svg}
\usepackage{graphicx}%
\usepackage{multirow}%
\usepackage{amsmath,amssymb,amsfonts}%
\usepackage{amsthm}%
\usepackage[title]{appendix}%
\usepackage{xcolor}%
\usepackage{textcomp}%
\usepackage{manyfoot}%
\usepackage{booktabs}%
\usepackage{algorithm}%
\usepackage{algorithmicx}%
\usepackage{algpseudocode}%
\usepackage{listings}%
\usepackage{gensymb}
\usepackage[nolist]{acronym}
\usepackage{hyperref}
\usepackage{url}

\begin{acronym}
\acro{NeRFs}{Neural Radiance Fields}
\acro{CNN}{Convolutional Neural Network}
\acro{MLP}{Multilayer Perceptron}
\acro{XMPI}{X-ray Multi-Projection Imaging}
\acro{DL}{Deep Learning}
\acro{FSC}{Fourier Shell Correlation}
\acro{SSIM}{Structural Similarity Index Measure}
\acro{3D}{three Dimensional}
\acro{XFELs}{X-ray Free-Electron Lasers}
\acro{STRT}{Super Time-Resolved Tomography}
\acro{MBIR}{Model-Based Iterative Reconstruction}
\acro{MSE}{Mean Squared Error}
\end{acronym}
\setcitestyle{super,comma}

\begin{document}

\title[Article Title]{Super time-resolved tomography }
\author*[1]{\fnm{Zhe} \sur{Hu}}\email{zhe.hu@sljus.lu.se}
\author[1]{\fnm{Kalle} \sur{Josefsson}}
\author[1]{\fnm{Zisheng} \sur{Yao}}
\author[2]{\fnm{Francisco} \sur{García-Moreno}}
\author[3]{\fnm{Malgorzata} \sur{Makowska}}
\author[1]{\fnm{Yuhe} \sur{Zhang}}
\author[1]{\fnm{Pablo} \sur{Villanueva-Perez}}
\affil[1]{\orgdiv{Synchrotron Radiation Research and NanoLund}, \orgname{Lund University}, \orgaddress{\city{Lund}, \postcode{22100}, \country{Sweden}}}
\affil[2]{\orgdiv{Institute of Applied Materials}, \orgname{Helmholtz-Zentrum Berlin}, \orgaddress{\city{Berlin}, \postcode{14109}, \country{Germany}}}
\affil[3]{\orgdiv{Laboratory of Nuclear Materials and Laboratory of Synchrotron Radiation and Femtochemistry}, \orgname{Paul Scherrer Institut}, \orgaddress{\city{Villigen}, \postcode{5232}, \country{Switzerland}}}

\abstract{Understanding \ac{3D} fundamental processes is crucial for academic and industrial applications. 
Nowadays, X-ray time-resolved tomography, or tomoscopy, is a leading technique for in-situ and operando 4D (\ac{3D}+time) characterization.
Despite its ability to achieve 1000 tomograms per second at large-scale X-ray facilities, 
its applicability is limited by the centrifugal forces exerted on samples and the challenges of developing suitable environments for such high-speed studies.
Here, we introduce \ac{STRT}, an approach that has the potential to enhance the temporal resolution of tomoscopy by at least an order of magnitude while preserving spatial resolution.
\ac{STRT} exploits a 4D \ac{DL} reconstruction algorithm to produce high-fidelity \ac{3D} reconstructions at each time point, retrieved from a significantly reduced angular range of a few degrees compared to the 0–180$\degree$ of traditional tomoscopy.
Thus, \ac{STRT} enhances the temporal resolution compared to tomoscopy by a factor equal to the ratio between 180$\degree$ and the angular ranges used by \ac{STRT}.
In this work, we validate the 4D capabilities of \ac{STRT} through simulations and experiments on droplet collision simulations and additive manufacturing processes.
We anticipate that \ac{STRT} will significantly expand the capabilities of 4D X-ray imaging, enabling previously unattainable studies in both academic and industrial contexts, such as materials formation and mechanical testing.
}

\maketitle

\section{Introduction}\label{sec1}

X-ray computed tomography (CT) is an essential tool across a plethora of fields in academia and industry for non-destructive study of three-dimensional (\ac{3D}) structures~\cite{withers2021x, villarraga-gomez_x-ray_2019,rawson_x-ray_2020}. 
CT works by collecting a series of 2D images, or radiographs, as the sample rotates relative to the X-ray source, covering an angular range between at least 0-180$\degree$~\cite{Cormack1964, Hounsfield1973}.
These radiographs are then combined using computing algorithms to reconstruct a \ac{3D} volume~\cite{KakSlaney2001}. 
The high flux provided by large-scale X-ray facilities, like diffraction-limited storage rings~\cite{eriksson_diffraction-limited_2014, white2021commissioning, Calvey:2024dyk} and \ac{XFELs}~\cite{Emma2010LCLS, Huang2012SACLA, Kang2017PALXFEL, Prat2020SwissFEL, Decking2020EuXFEL}, opens the possibility to study not only matter in \ac{3D} but also to resolve their evolution in 4D (\ac{3D}+time)~\cite{pandey2020time,garcia-moreno_time-resolved_2018}. 
Specifically, time-resolved X-ray tomography, also known as tomoscopy, has revolutionized our understanding of dynamic systems through \textit{in situ}, \textit{operando}, and \textit{in vivo} \ac{3D} characterization~\cite{garciamoreno_tomoscopy_2021, dos2014vivo}. 
The demand for tomoscopy stems from the critical need for non-destructive real-time analysis of internal structures and processes in various disciplines, such as fractures in solids~\cite{kumar_strength_2016, xu_different_2020}, manufacture and assembly of products~\cite{khosravani_use_2020}, fast biological processes~\cite{hansen_multifocal_2021, truong_high-contrast_2020}, and product degradation in service~\cite{shan_fatigue_2022}. 
For instance, understanding microstructural changes during manufacturing processes, such as alloy solidification, metal processing~\cite{kamm2025operando}, and mechanical deformation~\cite{garcia-moreno_xray_2023}, is critical for developing materials with enhanced properties. 
Traditional methods often fail to accurately capture these dynamic processes at the relevant spatiotemporal scale in a non-destructive manner, highlighting the importance of tomoscopy as a crucial tool for studying such \ac{3D} processes. 

Although tomoscopy is a well-established 4D tool, it faces challenges in improving its temporal resolution~\cite{holmes2023digital, gao2024dynamic}.
Specifically, it is crucial when studying dynamic processes to optimize temporal resolution without significantly compromising other parameters such as spatial resolution, field of view, and the total possible acquisition period. 
The current approach to increase the temporal resolution in tomoscopy involves increasing the rotation speed. 
The fastest tomoscopy experiments use high-speed rotation stages that enable rotations up to 500 Hz, which corresponds to one thousand tomograms per second~\cite{garciamoreno_tomoscopy_2021}. 
These acquisition speeds offer the opportunity to study processes such as alloy casting and sparkler burning~\cite{garcia-moreno_xray_2023}. 
However, these high-speed rotations induce centrifugal forces that are orders of magnitude times greater than the gravitational acceleration, potentially altering the studied dynamics.
In addition, they complicate the fabrication of sample environments for in situ and operando characterization.
Therefore, such fast rotations extremely limit the applicability of tomoscopy.
An alternative approach for studying processes in \ac{3D} is \ac{XMPI}, which can produce rotation-free \ac{3D} movies~\cite{villanueva-perez_hard_2018, bellucci_development_2024} at framerates from kHz and beyond~\cite{Asimakopoulou24kHzXMPI,villanueva-perez_megahertz_2023}. However, \ac{XMPI} suffers from sparse angular acquisitions, which may hinder the quality of the retrieved \ac{3D} data or its applicability to complex dynamics. 

Parallel to advances in 4D acquisition methods, there have been significant improvements in computational algorithms for image reconstruction.
Historically, reconstruction algorithms can be categorized by the dimensionality of the solutions: from 2D slices to full \ac{3D} volumetric reconstructions, and, more recently, to dynamic 4D approaches that include the time dimension~\cite{zhang_4d-onix_2024}.
The most common reconstruction approaches for tomoscopy rely on reconstructing individual 2D slices for each time point using: i) analytical solutions such as filtered back projection (FBP)~\cite{ziegler_noise_2007}, ii) iterative approaches~\cite{ning_x-ray_2014, fahimian_radiation_2013, lu_iterative_2023}, iii) compressed sensing~\cite{cuadros_coded_2017, yu_spectral_2016, bao_convolutional_2019}, and iv) \ac{DL}~\cite{liu_tomogan_2020, yang_tomographic_2020, pelt_improving_2018}.
\ac{DL} approaches have significantly improved the reconstruction quality of 2D slices, offering solutions to challenges such as sparse-view reconstruction and noise mitigation.
Moreover, \ac{DL} approaches have demonstrated the opportunity to provide direct \ac{3D} reconstructions for each time point, eliminating the need to stack 2D slices to retrieve the \ac{3D} volume.
Examples of these approaches include Convolutional Neural Networks (CNNs) with regular grids or continuous function representations~\cite{10.1007/978-3-030-61598-7_12, 10.1007/978-3-031-43999-5_2, shang2023stereo}.
However, CNN-based methods can be computationally expensive, limiting the volume that can be reconstructed in \ac{3D}.
More recent \ac{DL} approaches based on implicit representations such as \ac{NeRFs}~\cite{zhang_onix_2023, ruckert2022neat, zou2024pa} have shown the ability to efficiently reconstruct large \ac{3D} volumes with X-ray imaging.
When it comes to 4D reconstruction, there is a limited selection of algorithms that can be applied to tomoscopy.

Among the existing 4D reconstruction approaches, one prominent 4D approach is the use of \ac{MBIR} techniques~\cite{liu2014model} and, more recently, DYRECT~\cite{Goethals2024_dyrect}, such iterative approaches leverage temporal and spatial priors to enhance the reconstruction of dynamic objects. By using compressed sensing principles and sparsity constraints, these techniques can reconstruct high-quality 4D images from highly limited projection data~\cite{gao2024dynamic}. Despite their advantages, such iterative methods often face challenges in computational efficiency and robustness, especially in scenarios with complex motion patterns or highly dynamic samples.
In recent years, 4D-\ac{DL} computer vision approaches have shown the possibility of capturing complex 4D scenes in a computationally efficient manner, such as K-planes~\cite{fridovich2023k}, D-NeRF~\cite{pumarola2021dnerf}, and Hexplane~\cite{cao2023hexplane}.
Despite their potential, only a few of these algorithms have recently been adopted by the X-ray community, such as 4D-ONIX~\cite{zhang_4d-onix_2024}, designed explicitly for \ac{XMPI}.
These 4D X-ray approaches~\cite{zhang_4d-onix_2024, zheng2023ultrasparse} offer an opportunity to solve, tomoscopy directly in its intrinsic 4D space, thereby i) simplifying the reconstruction process, ii) imposing spatiotemporal consistency, and iii) offering opportunities to include 4D physical constraints. 
Thus, the combination of such algorithms together with state-of-the-art tomoscopy acquisition systems offer an opportunity to push the spatiotemporal resolution of 4D X-ray imaging.

In this paper, we present \ac{STRT}, a novel approach that enhances the temporal resolution of tomoscopy while preserving the spatial resolution.
This is achieved by reducing the angular range required to retrieve a \ac{3D} reconstruction for each temporal instance of the studied dynamics. 
\ac{STRT} exploits a deep learning approach to produce high-quality 4D reconstructions of dynamic objects using very few projections or radiographs within a narrower angular range than the current angular range required by tomoscopy: 0-180$\degree$.
This results in a temporal resolution enhancement equal to the ratio between the angular range used by \ac{STRT} and 0-180$\degree$.
We demonstrate with simulated and experimental data that \ac{STRT} can improve the temporal resolution by at least one order of magnitude compared to conventional tomoscopic experiments while preserving the spatial resolution. Such improvement stems from the possibility of reconstructing tomoscopic data in its intrinsic dimension (4D) and sharing features across space and time.
Consequently, we envision \ac{STRT} as an enabling tool that expands the possibilities of tomoscopy, not only by accessing academic and industrial processes previously inaccessible due to centrifugal forces or limited sample environments at large-scale X-ray facilities but also by enhancing the temporal resolution of laboratory equipment to bring tomoscopy into standard lab applications.

\section{Results}\label{sec2}

\subsection{Super time-resolved tomography}\label{subsec2.1}
\begin{figure}[htb!]
\centering
\includegraphics[width=0.95\textwidth]{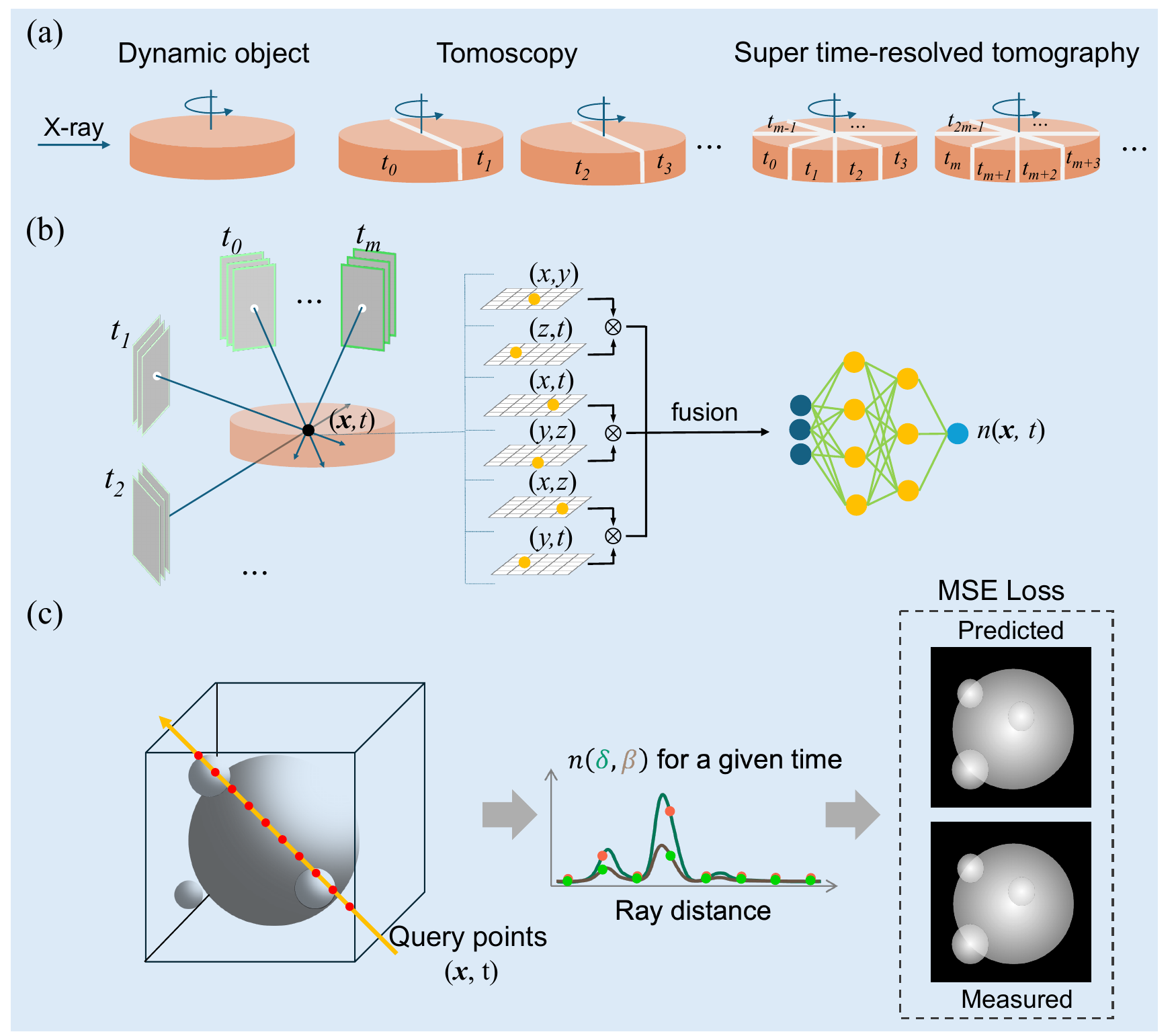}
 \caption{\ac{STRT} concept. (a) Data acquisition of \ac{STRT} vs. tomoscopy. Tomoscopy requires a full 180-degree scan to capture a single time point, while \ac{STRT} reduces the required scan angle to much less than $180\degree$. (b)  X-Hexplane tensorial model. X-Hexplane contains six feature planes spanning each pair of coordinate axes (e.g., XY, ZT).  Points in spacetime are projected to each plane. The features extracted from the six planes are fused and sent to a tiny \ac{MLP} to get $n(\textbf{x},t)$ at a specific spatiotemporal point. (c) X-Hexplane rendering and cost function. 2D projections from a given experiment angle and time point can be rendered by integrating $n(x,t)$ along the ray direction at that specific time point. The parameters of X-Hexplane are updated by minimizing the \ac{MSE} loss between the predicted results and the measured projections.}\label{STRT concept}
\end{figure}

State-of-the-art tomoscopy provides 4D information by reconstructing each time instance independently of a dynamical process from a complete series of radiographs acquired over an angular range of 0–180$\degree$, see Fig.~\ref{STRT concept}(a). This acquisition approach, including incremental reconstruction method~\cite{garcia2019using}, which shifts the start of each 180$\degree$ range by one projection, derives from conventional tomoscopy methods that reconstruct 2D slices and stack them into a \ac{3D} volume. 
In contrast, intrinsic 4D reconstruction approaches offer opportunities to increase temporal resolution by relaxing the acquisition protocols, exploiting 4D consistencies in the data, and simplifying both the reconstruction and analysis process.

Our approach, christened \ac{STRT}, enhances the temporal resolution of tomoscopy by utilizing a smaller angular range per time point compared to tomoscopy.
For this angular range, the sample is assumed static, and the corresponding projections are labeled with the same timestamp for reconstruction purposes.
The acquisition concept for \ac{STRT} is depicted in Fig.~\ref{STRT concept}(a).
\ac{STRT} uses a continuous rotation for the acquisition as done with tomoscopy, but each time point is defined over a smaller angular range, i.e., less than 180$\degree$ rotation.
This provides temporal enhancement that is proportional to the ratio between the angular range used in \ac{STRT} and 0-180$\degree$ used in tomoscopy experiments.
To enhance the temporal resolution while preserving the spatial resolution, \ac{STRT} exploits a 4D reconstruction framework that shares information between space and time throughout the acquisition, which alleviates the angular range requirement. 
In contrast, state-of-the-art CT reconstruction methods provide only 2D reconstructed slices for a specific time point.

Our 4D \ac{DL} reconstruction framework, X-Hexplane, is based on Hexplane~\cite{cao2023hexplane}, which utilizes a tensorial data structure to store 4D dynamic sample data~\cite{chen_tensorf_2022} and uses as input the recorded radiographs as a function of angle and timestamp. 
Unlike Hexplane, which is designed for visible light, we applied a physics-based forward X-ray propagation model to describe the interaction between matter and X-rays. 
Our forward model is based on the projection approximation (weak scattering)~\cite{paganin2006coherent}, where secondary scattering caused by X-ray photons is ignored.
This can be extended beyond the weak-scattering approximation by using, e.g., multi-slice methods~\cite{paganin2006coherent}. 
X-Hexplane describes a 4D dynamic sample in terms of the index of refraction at each spatial-temporal point $n({\bf x},t)$.
The index of refraction quantifies the X-ray interaction with matter and is represented as $n = 1 - \delta + i\beta$~\cite{paganin2006coherent}, where i) the real part $\delta$ captures the elastic interactions that result in wave phase shifts, offering high sensitivity to structural variations even in low-Z materials via the so-called phase contrast~\cite{ park_quantitative_2018}, and ii) the imaginary part $\beta$ relates to the inelastic interactions such as X-ray photon absorption and is related to the attenuation contrast~\cite{chen2012advances, withers2021x}. 

To provide a 4D computationally efficient reconstruction, X-Hexplane represents a dynamic process as an explicit voxel grid of features and reduces memory consumption by tensorial factorization~\cite{chen_tensorf_2022, cao2023hexplane}, as depicted in Fig.~\ref{STRT concept}(b). 
Specifically, X-Hexplane decomposes a 4D spacetime grid into six feature planes spanning each pair of coordinate axes, (XY, ZT), (XZ, YT), (XT, YZ). 
First, X-rays are marched from each pixel of the captured images, and points are randomly sampled along the rays.  
Each ($\textbf{x}$, $t$) sampled point on the left-hand side of Fig.~\ref{STRT concept}(b) is projected onto the six feature planes. Bilinear interpolation is applied to extract six corresponding feature vectors, which are then aggregated to form a fused feature vector. Subsequently, a lightweight \ac{MLP}, a simple feedforward neural network composed of fully connected layers, is employed to regress the fused vector and predict the value of the index of refraction at any given spatiotemporal point, $n(\textbf{x},t)$. 
See the Methods section for a more detailed description of the X-Hexplane algorithm. 
Once a differentiable description of the $n(\textbf{x},t)$ as a function of the input projections is provided, we can generate projections along any X-ray propagation direction and time by integrating the point values along the X-ray propagation direction using the projection approximation. 
In this work, a fixed distance from the detector and parallel beam geometry are used to compute the projections. 
Nonetheless, by modifying the geometric expression, cone-beam or fan-beam can also be implemented to account for different experimental configurations. 
Finally, to optimize X-Hexplane, we minimize an \ac{MSE} cost function between the predicted projections with X-Hexplane at a given time and the corresponding measured projections at the same time point.
This minimization process depicted in Fig.~\ref{STRT concept}(c) can be seen as an extension to 4D of current minimization approaches used in iterative reconstructions such as algebraic reconstruction techniques for 2D slice reconstructions\cite{ning_x-ray_2014}. 

To sum up, X-Hexplane provides a 4D X-ray reconstruction framework that enables \ac{STRT} by i) incorporating the physics of X-ray propagation into the model, ii) using a tensorial representation of the dynamics process to reduce memory footprint, and iii) sharing the features over space and time to solve the sparse view problems.

\subsection{STRT performance evaluation}\label{subsec2.2}
To evaluate and demonstrate the potential of the proposed X-Hexplane, we required accurate ground truth data. 
Therefore, we utilized simulated 4D data and standard tomoscopy experiments instead of \ac{STRT} experiments. 
The ground truth for the former was derived from the 4D simulated data, while for the latter, it was obtained from standard tomoscopy reconstructions at each time point using TomoPy~\cite{gursoy_tomopy_2014}.

To simulate an \ac{STRT} experiment from tomoscopy data, we developed a specialized data extraction that uses different continuous angular ranges for each time point.
Specifically, for each time point in \ac{STRT}, we selected projections that span a fixed angular range within a 360$\degree$ rotation to ensure continuous data acquisition over time. 
For instance, at time point $t_0$, projections covered an angular range from 0$\degree$ to $\theta\degree$, extracted from the first 360$\degree$ rotation of the tomoscopy data.
At time point $t_1$, projections covered $\theta\degree$ to 2$\theta\degree$, extracted from the second 360$\degree$ rotation of the tomoscopy data, and so on. 
It should be noticed that we extracted each time point from a 360$\degree$ rotation instead of the typical 180$\degree$ rotation used in tomoscopy.
This approach was chosen to avoid 4D inconsistencies that arise when combining $180^\circ$ data due to the flipping requirement for two consecutive time points, which can hinder the quality of X-Hexplane reconstructions. 
We emphasize that this data selection does not reflect a requirement of the \ac{STRT} acquisition scheme itself but rather a constraint imposed by using tomoscopy data as a ground-truth reference.
It should be noted that an \ac{STRT} experiment would use all the continuous frames acquired through a continuous rotation. 
Then, the reconstruction method would be applied directly to the recorded projections without any data selection or even requiring a full or half-rotation dataset.

Table~\ref{tab1} reports the different angular ranges and projections data used in our validation experiments for tomoscopy and \ac{STRT}, together with the temporal enhancement (TE) between \ac{STRT} and tomoscopy.
To assess the resolution for a given temporal enhancement, we compared the quality of the \ac{STRT} reconstructions with the ground truth as a function of the angular range employing the \ac{FSC}~\cite{van_heel_fourier_2005}.
The \ac{FSC} calculates the normalized cross correlation between the reconstructions and the ground truth in frequency space over shells. 
Therefore, it provides a robust metric for comparing the spatial frequency content of the two results, ensuring a thorough evaluation of the neural network's performance for each time point. 
Here, we used \ac{FSC} with the half-bit threshold criterion to determine the achievable resolution~\cite{van_heel_fourier_2005} and the results for each process are summarized in Table~\ref{tab1}.

In the following sections, we describe in detail our validation results for the simulated and experimental data.
\begin{table}[ht]
\begin{tabular}{c| c c| c l c c}
\Xhline{2pt}
   & \multicolumn{2}{c|}{\textbf{Tomoscopy}}   & \multicolumn{3}{c}{\textbf{STRT}}\\
   \hline
Process    &\makecell{No. of \\timesteps} & \makecell{No. of \\projs} & \makecell{No. of \\timesteps} & \makecell{No. of projs\\(Angular range)} & \makecell{Temporal \\enhancement} & \makecell{\ac{FSC} \\Mean (Std)}\\
\hline
\multirow{3}{*}{\makecell{Droplet collision}}   &\multirow{3}{*}{\makecell{75}}  &\multirow{3}{*}{\makecell{180}}   &\multirow{3}{*}{\makecell{75}}   &~~~~~3  (3.0$\degree$)        & ~60$\times$ & 4.1 (1.8) \\
                                                                                     &  &   &     &~~~~~9  (9.0$\degree$)       & ~20$\times$  & 3.0 (0.4)\\
                                                                                     &  &   &     &~~ ~~18 (18.0$\degree$)       & ~10$\times$ & 2.8 (0.3) \\
\hline
\multirow{3}{*}{\makecell{Additive\\ manufacturing}}   &\multirow{3}{*}{\makecell{200}} &\multirow{3}{*}{\makecell{200}}   &\multirow{3}{*}{\makecell{200}}   &~~~~~1  (0.9$\degree$)        & ~~200$\times$  & 5.7 (0.8)\\
                                                                                     &   &   &    &~~~~~10  (9.0$\degree$)       & ~20$\times$  & 3.7 (0.6)\\
                                                                                     &   &   &    &~~~ ~20 (18.0$\degree$)       & ~10$\times$  & 3.2 (0.5)\\
\Xhline{2pt}
\end{tabular}
\caption{Result evaluation for the simulated and experimental datasets as a function of the temporal enhancement. Columns represent the number of timesteps (No. of timesteps), the number of projections (No. of projs) at each timestep for tomoscopy and \ac{STRT}, and the temporal resolution enhancement of \ac{STRT} compared to tomoscopy. The last column shows the mean and standard deviation (Std) of \ac{FSC} results over the entire time sequence as a function of the tomoscopy voxel size.}\label{tab1}%
\end{table}
\subsection{Simulation results}\label{subsec2.3}

We first assessed the performance of \ac{STRT} using 4D simulated datasets describing a coalescence process as a result of a binary droplet collision~\cite{adam1968collision, qian1997regimes, planchette2012onset}. This 4D process was modeled using Navier-Stokes Cahn-Hilliard equations~\cite{zimmermann2019calculation, zhang_4d-onix_2024}. 
To be consistent with a potential \ac{STRT} experiment on such a system, we assumed that the X-ray source was continuously rotated instead of the sample~\cite{fischer2010ultra, lau2018ultrafast}. 
For each of the 75 simulated time points in this process, projections with $128 \times 128$ pixels were generated. The modeled data was used to simulate an \ac{STRT} experiment. 
We extracted continuous subsets of projections that cover angular ranges of 3$\degree$, 9$\degree$, and 18$\degree$ per time point for training, which corresponded to 60$\times$, 20$\times$, and 10$\times$ temporal enhancement, respectively.

\begin{figure}[htbp!]
\centering
\includegraphics[width=0.95\textwidth]{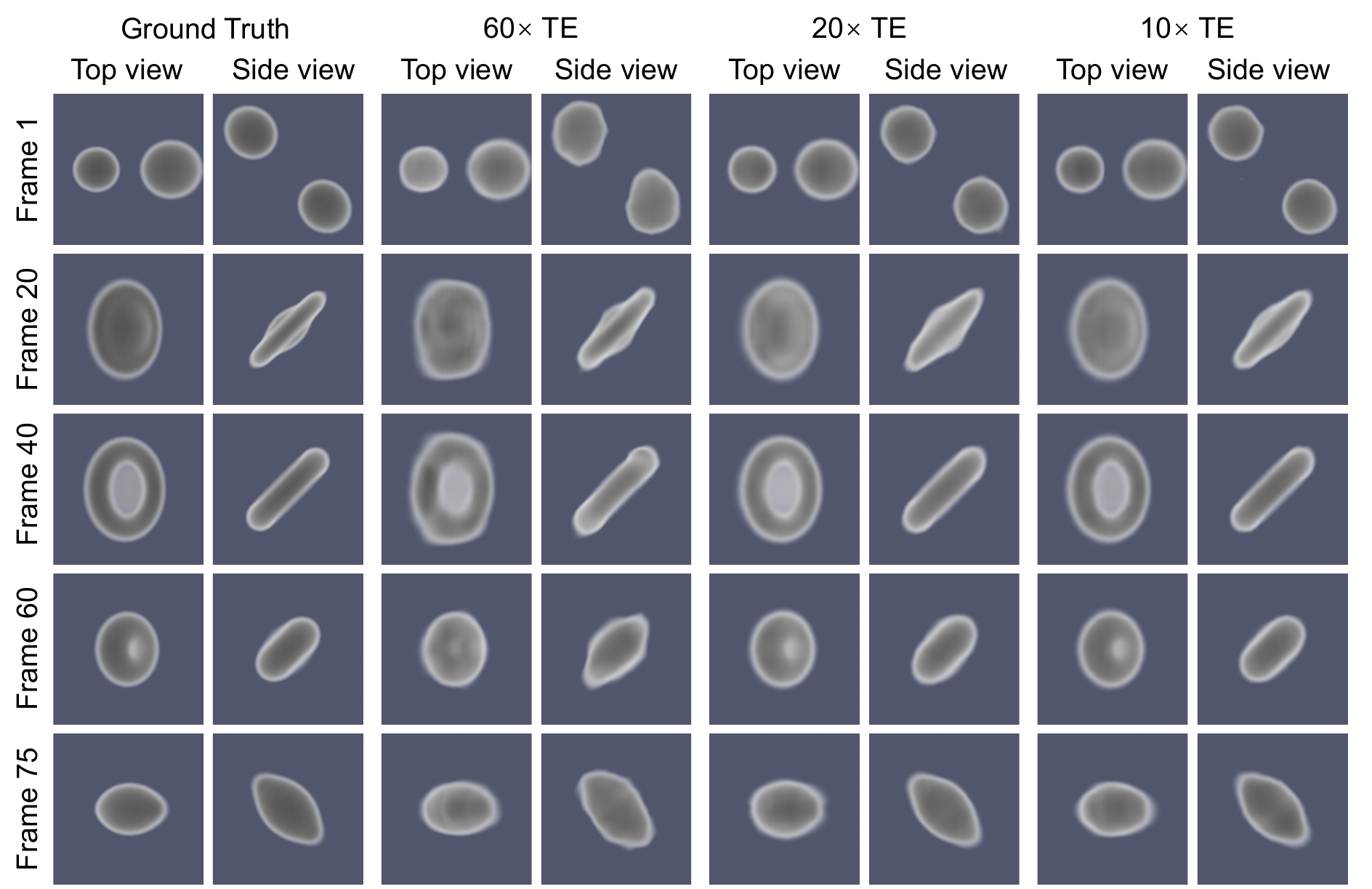}
 \caption{\ac{STRT} reconstructions for droplet collisions compared to the ground truth. The left column depicts the 4D ground truth for different time points (frames) from two orthogonal views: one along the projection axis (side view) and one perpendicular to this plane (top view). The other columns show the corresponding \ac{STRT} reconstructions for 60$\times$, 20$\times$, and 10$\times$ temporal enhancement (TE).\label{simulation results}
}
\end{figure}

\begin{figure}[htbp!]
\centering
\includegraphics[width=0.95\textwidth]{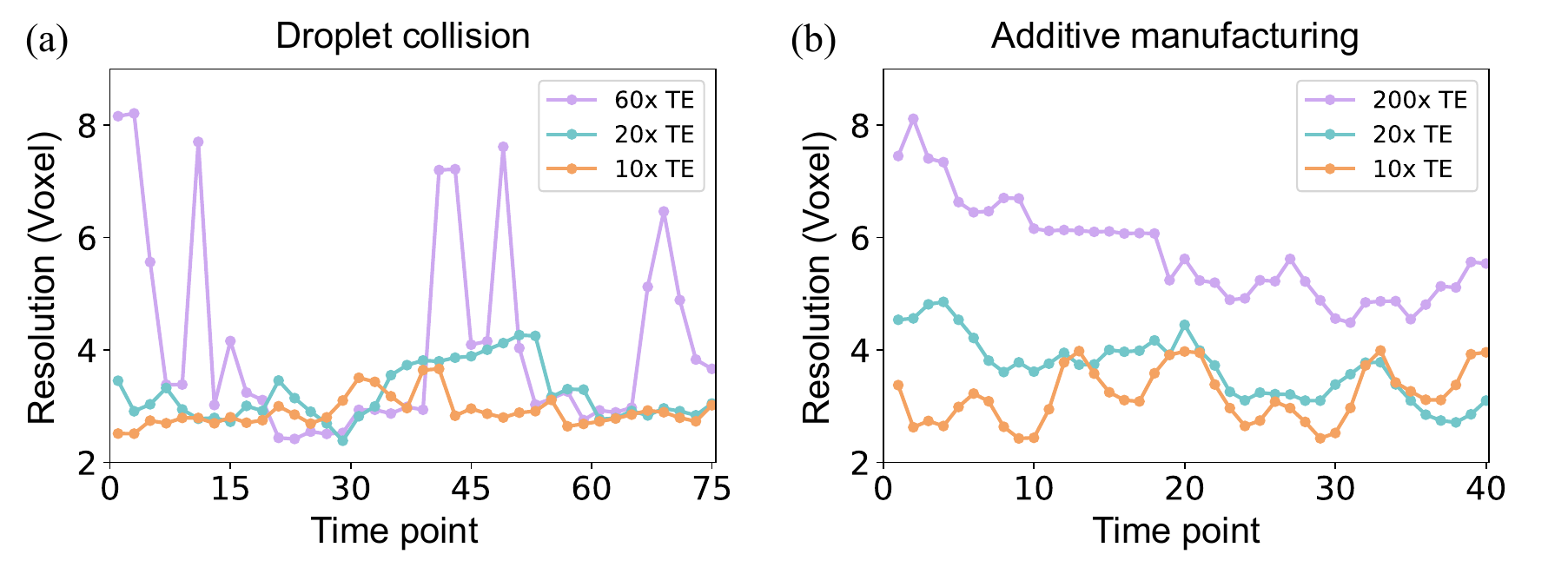}
 \caption{Spatial resolution determined by \ac{FSC} as a function of time. (a) illustrates the resolution changes during droplet collision with temporal enhancements (TE) of 60$\times$, 20$\times$, and 10$\times$. (b) presents the evolving revolution in additive manufacturing over time with temporal enhancements of 200$\times$, 20$\times$, and 10$\times$.}\label{resolution map}
\end{figure}

Fig.~\ref{simulation results} presents the ground truth and reconstruction volumes as a function of the temporal enhancement from two orthogonal views at specific time points. A side view along the projection plane and a top view perpendicular to it were selected. The top view is particularly challenging to reconstruct, as it aligns with the rotation axis and is perpendicular to the incident X-rays.
Additionally, the Supplementary Material provides a video that shows a comparison between ground truth and \ac{STRT} reconstruction results as a function of the temporal enhancement. 
The quantitative reconstruction evaluation via the \ac{FSC} is summarized in Table~\ref{tab1}.
The \ac{FSC} as a function of time for the binary droplet collisions is shown in Fig.~\ref{resolution map}(a).
For the reconstructions with an angular range of just 3$\degree$, i.e., a temporal enhancement factor of $60$, the reconstructed \ac{3D} movie achieved an average resolution of 4.1 voxels, only two times worse than the Nyquist-limited resolution of two voxels.  
In the 20$\times$ and 10$\times$ cases, the results successfully capture both shapes and dynamics across all frames, achieving an average resolution of less than 3 voxels almost compatible for all the time points with the Nyquist resolution.

\subsection{Additive manufacturing results}\label{subsec2.4}
X-ray tomoscopy experiments on laser-based powder bed fusion (LPBF) of iron oxide doped alumina were performed at the TOMCAT beamline of the Swiss Light Source. The tomoscopy presented in this work was performed while scanning the sample with a speed of 1 mm/s using a green (532 nm) pulsed laser (1.5 ns pulse duration) with power set to 20 W. The observed process was remelting of the previously deposited layer, meaning that in this case, no new powder layer was deposited in the sample and the laser interacted directly with the solidified ceramics. The detailed description of the used material and the setup for the operando tomography during the LPBF process can be found in Ref.~\cite{makowska_2023_LPBF}.
The studied LPBF process consists of two stages. 
The first involves a powder layer deposition. In the second stage, laser scanning a selected region of the sample induces localized melting of the powder, which is followed by the cooling down and solidification of the molten material. By adding consecutive multiple layers, a stable, three-dimensional structure is obtained. The morphology and microstructure of LPBF-manufactured materials are strongly linked to the dynamics of the melt pool. To track the behavior of the molten volume and the solidification processes, tomoscopy measurements with an acquisition rate of 100 tomograms per second were recorded, each with 200 projections over 180$\degree$.
This high temporal resolution was achieved using the high polychromatic flux provided by the TOMCAT superbending magnet source. 
Due to the limited temporal (broad spectrum) and spatial (large source size) coherence of the TOMCAT source, limited phase contrast was observed in the projections. 
Thus, we performed phase-retrieval using transport of intensity equations for homogeneous objects~\cite{Paganin2002TIE}, which resulted in a single reconstructed component.
Our X-Hexplane architecture was, therefore, tailored for this experiment to be sensitive only to the phase-retrieved component of the index of refraction. 
The projections or radiographs were acquired with the GigaFRoST camera system~\cite{mokso_gigafrost_2017}. 
The camera system provided an effective pixel size of 2.75 µm in this experiment, with radiographs measuring $160 \times 1056$ pixels for additive manufacturing. To enhance visualization and accelerate the reconstructions, the top part of the object was selected, and the width was scaled down (see Methods section), resulting in a final projection size of $70 \times 528$ pixels. Moreover, we selected the $200$ time points where the printing process took place. 
Datasets were prepared with different temporal enhancement of 200$\times$, 20$\times$, and 10$\times$,  corresponding to angular ranges of 0.9$\degree$, 9$\degree$, and 18$\degree$ per time point, respectively.
The specific details of the dataset processing are discussed in the Methods section.

\begin{figure}[htbp!]
\centering
\includegraphics[width=0.95\textwidth]{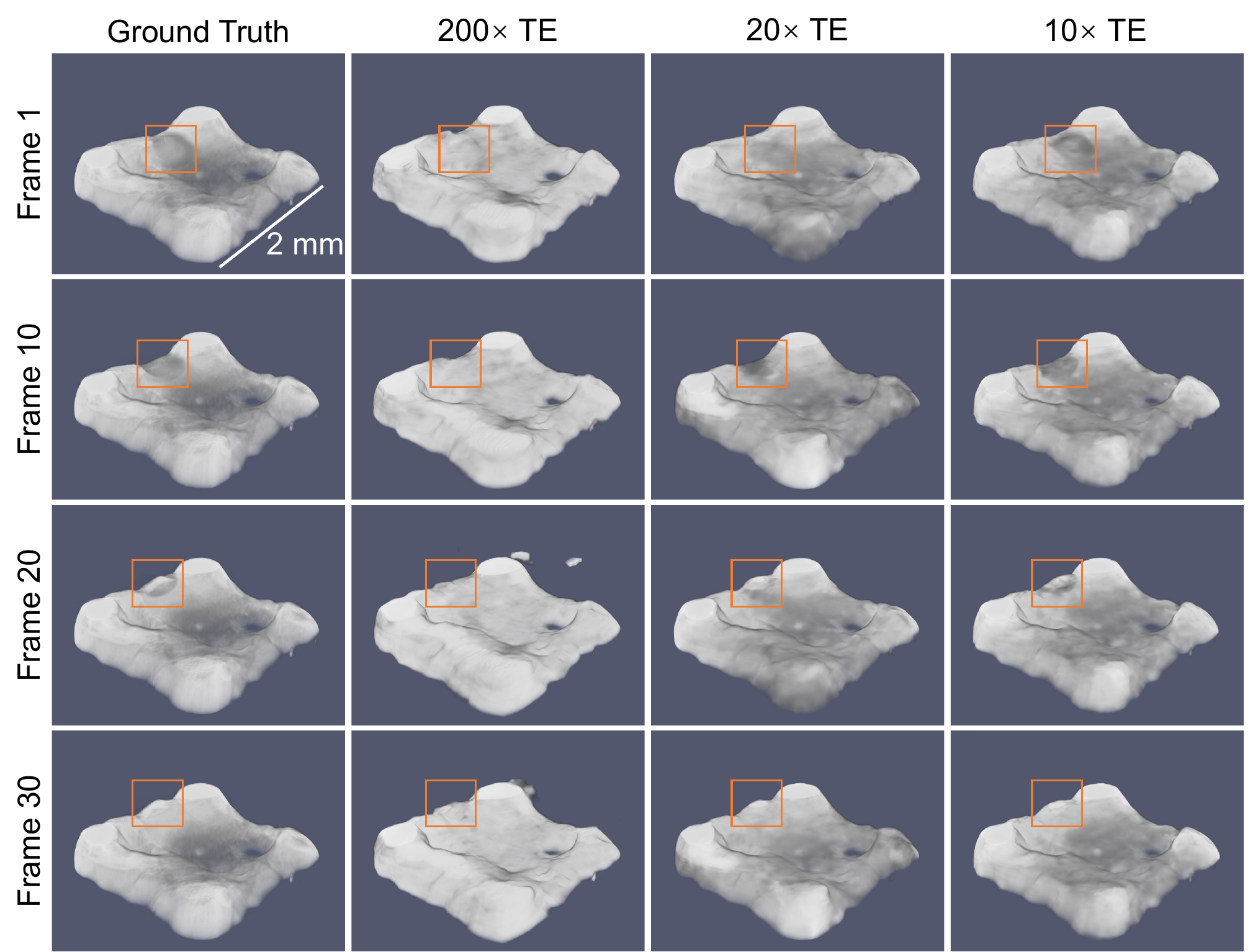}
 \caption{ \ac{STRT} reconstructions for additive manufacturing and ground truth. The ground truth and \ac{STRT} results for 200$\times$, 20$\times$, 10$\times$ temporal enhancement (TE) are displayed in separate columns, while different rows represent distinct time frames of the printing process. The dynamics of the printing processes are marked by solid squares.}\label{am results}
\end{figure}

Fig.~\ref{am results} presents the comparison between the ground truth obtained from full 180$\degree$ per time point and the corresponding \ac{STRT} reconstructions. 
The different rows in the figure depict the sample's state at various time points. A video comparing the reconstructions and the ground truth is provided in the Supplementary Material.
As it can be observed in Fig.~\ref{am results}, \ac{STRT} produces promising results even with only 1 projection per time point, effectively capturing the general shape of the sample. However, noticeable artifacts and distortions appear, particularly in finer structures in the area that is being printed (orange squares in Fig.~\ref{am results}).
The reconstruction corresponding to 20$\times$ temporal enhancement improved overall shape preservation but still struggles with accurately recovering the finer details of the printing process. With 10$\times$ temporal enhancement, both deformation stages were successfully reconstructed, indicating that an angular range of 18$\degree$ per time point is sufficient to capture the dynamic process. 
The mean and standard deviation of the FSC results are reported in Table~\ref{tab1}, and the FSC as a function in time is shown in Fig~\ref{resolution map}(b).
Such results confirm that the best results are achieved with 10$\times$ enhancement with a resolution of approximately three voxels compared to the ground truth.

\begin{figure}[htbp!]
\centering
\includegraphics[width=0.95\textwidth]{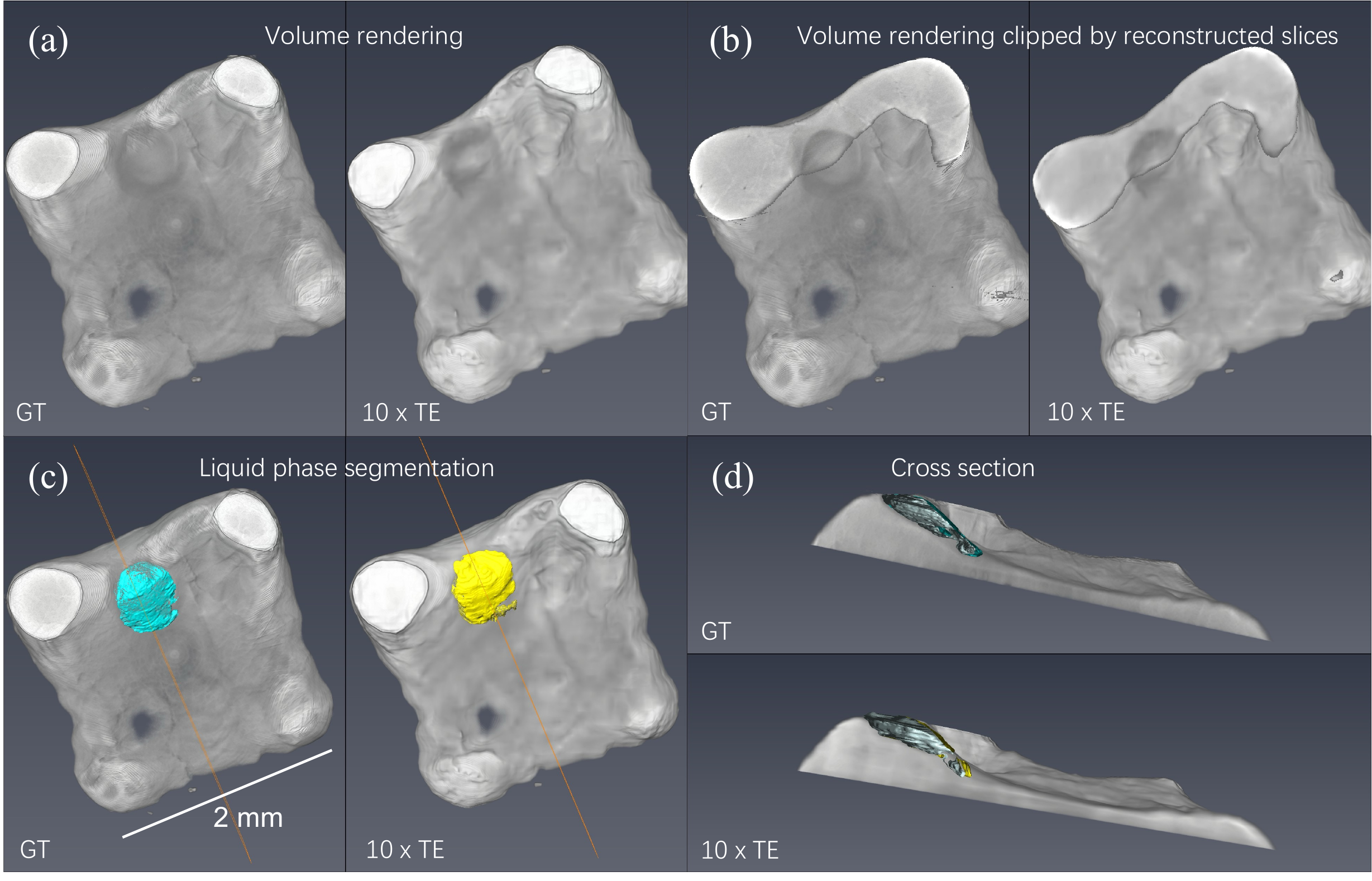}
 \caption{ Comparison of  the reconstruction performed with 10$\times$ temporal enhancement (TE) with the ground truth (a) volume rendering, (b) volume rendering clipped by slices reconstructed using corresponding approaches (ground truth and 10$\times$ TE), (c) results of material segmentation highlighting liquid surfaces, (d) cross sections along the orange line indicated in (c).}\label{AM segmentation}
\end{figure}

For further evaluation of the quality of the obtained reconstructions and to assess their applicability for the studies of the LPBF process, material segmentation was performed using standard tools available in the commercial software Thermo Scientific Avizo. The results are illustrated in Fig.\ref{AM segmentation}. Fig.\ref{AM segmentation} (a) and (b) show the comparison of the reconstructed volume obtained with 10$\times$ enhancement with the ground truth by volume rendering and the obtained tomographic slices, clipping the same volumes approximately through the middle of the melt pool depth. A simple image contrast adjustment allows an obvious distinction between the solid and liquid phases, which in turn facilitates the material segmentation. The liquid surface generated based on the segmentation is displayed in Fig.\ref{AM segmentation} c) (top view) and d) (cross section) and clearly shows the similar shape of the obtained melt pool in both cases. The volume of the molten phase obtained for the ground truth reconstruction was 1.4$\times10^6$ $\mu m^3$, and for 10$\times$ temporal enhancement was 1.6$\times10^6$ $\mu m^3$. This difference is within the possible deviation range due to the obtained spatial resolution. 

\section{Discussion}\label{sec3} 
\ac{STRT} has demonstrated the potential to significantly enhance the temporal resolution of tomoscopy in both simulated and experimental cases, achieving at least a tenfold improvement while preserving spatial resolution. 
The highest achievable temporal enhancement with \ac{STRT} while preserving the tomoscopy spatial resolution depends on the specific dynamics being studied. Thus, we discuss each test independently.

In the case of droplet collision, as shown in Fig.~\ref{resolution map}(a), intermediate collision states are more challenging to reconstruct across all temporal enhancement levels due to the increased complexity of the dynamics. 
The degradation in reconstruction quality is particularly evident at 60× enhancement, where the resolution is lower, and the reconstructed structures fail to capture the precise morphology of the droplets. 
This suboptimal performance can be attributed to the limited angular coverage for each time point, which is more dramatic for abrupt changes in transient states.
Specifically, for the 60$\times$ temporal enhancement, the total angular coverage through the \ac{STRT} is restricted to 75 $\times$ 3$\degree$ = 225$\degree$ while having the same number of time intervals as the 20$\times$ and 10$\times$ temporal enhancement cases. 
This limited information provided to the network makes the transient state even more challenging to reconstruct. 
In contrast, at 10$\times$ and 20$\times$ enhancement, the reconstructions successfully capture both fine and coarse structural details across all frames. In Fig.~\ref{resolution map}(a), the 10$\times$ temporal-enhancement curve appears more stable with fewer fluctuations, whereas the 20$\times$ enhancement curve shows greater variability at transient stages.
This demonstrates that the larger angular range per time point provides high-quality spatiotemporal reconstructions even in the presence of fast dynamics with respect to the acquisition framerate.

The best resolution reconstruction for the laser powder bed fusion tomoscopy is achieved at 10$\times$ enhancement but deteriorates significantly at 200$\times$ enhancement as depicted in Fig.~\ref{resolution map}(b). 
The 200$\times$ enhancement fails to accurately reconstruct the morphology of the laser-processed part and to provide a clear contrast between the liquid and solid phase, with an average resolution of 6.9 voxels over the first 10 time points. 
This degradation mainly arises from the extremely limited angular coverage per time point—only 0.9$\degree$ per time frame, leading to a total angular coverage of 200 $\times$ 0.9$\degree$ =180$\degree$ over the full 4D acquisition. 
The restricted angular information at individual time points and across the whole acquisition sequence hinders the algorithm’s ability to retrieve 4D shareable information effectively, particularly in regions with complex geometries and rapid dynamics. 
As a result, it is not possible to image structural defects such as cracks and pores appearing in the sample nor to analyze the dynamics of the molten phase since it cannot be distinguished from the solid phase. 
For the 20$\times$ case, although the reconstruction still struggles to precisely capture the phenomena occurring in the region interacting with the laser and therefore changing over time (with an average resolution of 4.2 voxels in the first 10 time frames), it successfully captures the overall shape of the manufactured alumina part across all frames, representing a significant improvement in \ac{FSC} results. 
Finally, the 10$\times$ enhancement results provide a better and more stable \ac{FSC} resolution over time. 
The improved reconstructions at 10$\times$ and 20$\times$ enhancement compared to the 200$\times$ are attributed to the larger angular range per time point, which provides more angular information with limited time points. 
This additional information enhances the algorithm's capability to reconstruct both fine and coarse structural features with greater accuracy. 
Such results demonstrate that \ac{STRT} successfully retrieves the dynamic process consisting of the 3D reconstructions for each time point depicted in Fig~\ref{am results}, providing a sufficient contrast between the observed phases (solid and liquid). This is clearly demonstrated in Fig. \ref{AM segmentation}, presenting the results of material segmentation. Both the shape and the volume of the melt pool evaluated from the reconstruction based on 10$\times$ temporal enhancement were in good agreement with the ground truth. It should be noted that in the case of the studied material, namely, alumina, the contrast between solid and liquid is inherently low. Therefore, this data set, though very relevant for real applications, is a particularly challenging case.

Although \ac{STRT} offers clear advancements over tomoscopy in additive manufacturing, the data selection and processing required to compare \ac{STRT} with tomoscopy as a ground truth introduces artifacts that limit the full potential of \ac{STRT}. 
For example, one should notice that the resolution curve in Fig.~\ref{resolution map}(b) for 10$\times$ enhancement exhibits a periodic pattern with a 20-time-step interval, which arises from the fact that projections are acquired from the same viewing angle of the sample every 360$\degree$ / 18$\degree$ = 20 time steps. 
These periodic peaks in the curve result from noise and inconsistencies between the alignment of projections when interpreting tomoscopy data as \ac{STRT}.
Another key effect that causes such inconsistencies is edge enhancement due to X-ray propagation.
Edge enhancement effects refer to the phenomenon where edges of structures appear more pronounced due to phase shifts in the transmitted wavefront, leading to interference effects that enhance contrast at sample boundary regions. 
When viewed from certain angles, one boundary of the object may overlap with another boundary region, enhancing this effect.
While phase retrieval methods~\cite{Paganin2002TIE} can remove the edge enhancement effect from certain projections, they can introduce errors in projections that contain overlap regions, leading to inconsistencies between different projections together with resolution loss. 
These inconsistencies can affect the accuracy of the \ac{STRT} reconstructions, particularly in regions with complex geometries or fine structural details. 
As such overlap is produced in a periodic manner due to the rotation, one can observe this periodicity in the peaks of the \ac{FSC} curve.

In spite of the \ac{STRT} improvements in temporal resolution compared to state-of-the-art methods, several challenges and limitations remain.
The algorithm relies on resolving features at each spacetime point, making it less effective for datasets with high noise levels or faint features. 
In such cases, the improvement in temporal resolution is constrained by the flux per projection, necessitating longer exposure times to reduce noise and sufficiently resolve relevant features within each projection or angular range.
Moreover, in high temporal enhancement \ac{STRT} experiments, dynamical processes with abrupt changes occurring within only a few projections or frames present significant challenges, as the limited angular information available for sharing across the time sequence is insufficient to accurately capture these rapid transitions.
Increasing the number of time points (framerate) while ensuring minimal sample changes between consecutive frames can further improve reconstruction quality at higher temporal enhancements. 
This approach requires a fast-framerate camera and a high-flux X-ray source to leverage the information-sharing capabilities of STRT and compensate for the missing angular views. 
Additionally, transferring knowledge across different experiments on the same or similar dynamical process can serve as an effective strategy to mitigate such an effect~\cite{zhang_4d-onix_2024}. 
By conducting multiple experiments and optimizing them collectively, the model can learn shared features across not only 4D but also the different experiments, leading to better convergence and improved accuracy in the reconstruction results.
Finally, incorporating the physics of the studied dynamics to further constrain the process can also enhance spatiotemporal knowledge sharing~\cite{raissi_physics-informed_2019, yao2025physics}, thereby mitigating limitations and improving reconstruction fidelity.

To conclude, we have introduced \ac{STRT}, a 4D X-ray imaging approach that enhances the temporal resolution of the state-of-the-art 4D X-ray imaging technique (tomoscopy) by exploiting a new acquisition approach and a self-supervised 4D \ac{DL} framework. 
Compared to tomoscopy, \ac{STRT} enhances the temporal resolution by at least an order of magnitude while preserving spatial resolution, addressing the limitations imposed by centrifugal forces and the challenges of developing suitable environments for 4D high-speed studies with X-ray imaging. 
By exploiting a 4D \ac{DL} reconstruction algorithm, \ac{STRT} achieves this enhancement through several key advancements: i) incorporating the physics of X-ray propagation into the model, ii) using a tensorial representation of the dynamics process to reduce memory footprint, and iii) sharing the features over time to solve the sparse view problems. 
We have demonstrated the capabilities of \ac{STRT} through proof-of-concept experiments on both simulated and experimental data, where high-quality reconstructions were achieved with a temporal resolution enhancement of at least a factor of 10 without compromising spatial resolution. 
The ability to capture these rapid phenomena with a tenfold increase in temporal resolution enables a more detailed understanding of transient dynamics, defect formation, and structural evolution in such systems. 
Specifically, in the case of the demonstrated example of the laser powder bed fusion studies, the increase of the temporal resolution will allow the scientific community to broaden the range of studied materials. 
In this process, the scanning speed of the laser determines the speed of the melt pool translation. 
In the case of ceramics, the scanning laser speed is typically low (several to several tens of mm per s), and therefore, tomoscopy with temporal resolution of 100 tps allowed to track the melt pool dynamics in alumina~\cite{makowska_2023_LPBF}. 
However, for metals, the laser speed is typically one or two orders of magnitude higher, and therefore conventional tomoscopy would not be possible to apply due to the need to increase the rotation speed. 
Thus, the possible gain in temporal resolution presented in this work can enable 4D imaging of LPBF of materials, which was not possible with previously used methodology. 
This type of experimental data will be crucial for understanding the physical phenomena determining the microstructure and quality of the additively manufactured materials but also will provide information necessary for validation and development of numerical models for additive manufacturing processes, which was highlighted in Ref.~\cite{MUTHER2025104756}.
To sum up, the significant enhancement in temporal resolution achieved with \ac{STRT} opens new possibilities to investigate industrially and scientifically relevant processes that were previously inaccessible with tomoscopy and other advanced 4D X-ray imaging methods.
\ac{STRT} will not only be impactful at large-scale X-ray facilities by enabling 4D imaging at rates exceeding 1000 tomograms per second but also enhancing 4D imaging using X-ray laboratory sources.

\section*{Methods}

\subsection*{X-Hexplane algorithm}
Here, we describe the implementation of X-Hexplane, a 4D reconstruction framework that builds upon the Hexplane framework~\cite{cao2023hexplane} to optimize memory usage and computational efficiency. 
X-Hexplane employs a grid-based representation with six planes, where low-rank tensors are embedded within each pixel of the planes. 
By representing dynamic samples as 4D tensors, X-Hexplane enables the factorization of these tensors into multiple low-rank components, significantly reducing memory requirements while maintaining high reconstruction accuracy. 
This structure allows for efficient modeling and reconstruction of spacetime data by discretizing the domain and capturing essential features through the embedded tensors. 
As described in TensoRF~\cite{chen_tensorf_2022}, \ac{3D} volume (V) can be decomposed as a sum of the outer products of vector-matrices:

\begin{equation}
V=\sum_{r=1}^{R_1}M_{r}^{x,y}\otimes v_{r}^{z}\otimes v_r^1+\sum_{r=1}^{R_{2}}M_{r}^{x,z}\otimes v_{r}^{y}\otimes v_{r}^{2}+\sum_{r=1}^{R_{3}}M_{r}^{y,z}\otimes v_{r}^{x}\otimes v_{r}^{3}~,\label{3drepr}
\end{equation}
where $\otimes$ is the outer product, $M_{r}^{x,y}\otimes v_{r}^{z}\otimes v_r^1$ is the component corresponding to different axes; $M_{r}^{x,y}\in R^{XY}$, $M_{r}^{x,z}\in R^{XZ}$, and $M_{r}^{z,y}\in R^{ZY}$  are matrices spanning the (\textit{X}, \textit{Y}), (\textit{X}, \textit{Z}),  (\textit{Y}, \textit{Z}) axes and $V_{r}^{x}\in R^{X}$, $V_{r}^{y}\in R^{Y}$, $V_{r}^{z}\in R^{Z}$, and $V_{r}^{i}\in R^{F}$ are vectors of feature. \textit{R}\textsubscript{1}, \textit{R}\textsubscript{2}, \textit{R}\textsubscript{3} are the number of low-rank components.
Factorization is known to reduce memory usage; however, factoring an independent \ac{3D} volume for each time step poses challenges due to sparse observations in the experimental configuration and the inability to share information across time points. To address this issue, the approach~\cite{cao2023hexplane} represents the \ac{3D} volume \textit{V}\textsubscript{\textit{t}}  as the weighted sum of a set of shared \ac{3D} basis volumes ${\hat{V}_i}$. 
\begin{equation}
V_{t}=\sum_{i=1}^{R_{t}}f(t)_i\cdot \hat{V_{i}}.\label{t_weighted}
\end{equation}
Replacing $v_r^1\cdot f^1(t)_r$ with a joint function of \textit{z} and \textit{t}, similar to the matrix spanning the \textit{X} and \textit{Y} axes, Eq.~\ref{3drepr} becomes:
\begin{equation}
V_{4D}=\sum_{r=1}^{R_1}M_r^{x,y}\otimes M_r^{z,t}\otimes v_r^1+\sum_{r=1}^{R_2}M_r^{x,z}\otimes M_r^{y,t}\otimes v_r^2+\sum_{r=1}^{R_3}M_r^{y,z}\otimes M_r^{x,t}\otimes v_r^3~,\label{4drepr}
\end{equation}
where $V_{4D}$ is the feature representation of 4D volume, and each $M_r^{a, b}\in{R^{A,B}}$ is a learned plane of features, $v_r^i$ are the vectors of feature. In this case, the spacetime complexity is reduced from \textit{O}(\textit{N\textsuperscript{3}T}) to \textit{O}(\textit{N\textsuperscript{2}F}), where $N$, $T$, and $F$ are the spacial resolution, the temporal resolution, and feature size, significantly reducing the memory footprint. 
X-rays are traced from each pixel of the projections along the direction in which the projections were acquired. Spacetime points are randomly selected along the X-rays and projected onto the six planes. 
Six corresponding tensors are then extracted using bilinear interpolation. These tensors are fused and passed to a \ac{MLP} to regress the $n(\textbf{x}, t)$ for each point. 

After calculating $n(\textbf{x}, t)$, 2D images or projections can be rendered using the projection approximation. The projection approximation can be described by:
\begin{equation}
\psi_{z_{exit}}=\psi_{z_0} exp\Big(-ik\int_{z_0}^{z_{exit}} (\delta[x,y;z]-i\beta [x,y;z])dz\Big)~,\label{wave}
\end{equation}
where  $\psi_{z_{0}}$ is the incoming wave on the sample, $\textit{k} = 2\pi/\lambda$  is the wavenumber which is inversely proportional to the wavelength ($\lambda$), $\delta$, $\beta$ are a function of the transverse coordinates \textit{x} and \textit{y} orthogonal to the propagation direction \textit{z}. This formulation enables the linear integration along the X-ray propagation direction.

X-Hexplane is optimized by \ac{MSE} loss between rendered and target images. To leverage the ill-defined sparse view problem, Total Variation (TV) loss is used on planes to force the spatial-temporal sparsity and continuity as a regularizer. The optimization cost function or loss is given by: 
\begin{equation}
L = \frac{1}{\lvert R \rvert} \sum_{r \in R} \lVert c(r) - \hat{c}(r) \rVert_2^2 + \lambda_{\text{reg}} L_{\text{reg}}.\label{loss}
\end{equation}
where \textit{R} is the set of all points and $\hat{c}(r)$ denotes the estimated phase or attenuation contrast; $L_{reg}$, $\lambda_{reg}$ are regularization and its weight. By minimizing the difference between the real and predicted projections, the network parameters are updated, allowing it to produce more accurate and realistic 4D reconstructions. 

In the presented studies, \ac{STRT} was tailored to generate a single channel of the index of refraction. However, its adaptability and flexibility make it applicable to a wide range of time-resolved imaging experiments across various imaging domains. For instance, it can be extended to techniques such as coherent diffraction imaging~\cite{miao_coherent_2012, schroer_coherent_2008} and phase-contrast imaging~\cite{cuche_digital_1999, park_quantitative_2018}, where the propagation model is explicitly known and can be easily incorporated into the X-ray imaging model of \ac{XMPI}~\cite{zhang_onix_2023,zhang_4d-onix_2024}. 

\subsection*{Network and training details}
For the implementation, we adopted grid sizes for the six planes that scale according to the dimensions of the input dataset. In the drop collision case, the grid size for both spatial and temporal dimensions was set to 64. For the additive manufacturing case, the grid size was increased to 128 to accommodate the higher complexity of the scenario. In both cases, the embedded low-rank tensors had a dimension of (48) for all grids, and the \ac{MLP} architecture consisted of 3 layers, with each layer containing 64 neurons. This configuration ensures a balance between computational efficiency and model performance.

\ac{STRT} was implemented using PyTorch 1.6.0 and Python 3.8.8. The optimization and training processes were conducted on an NVIDIA A100 GPU with 80 GB of RAM. The training time for 50,000 epochs was approximately 10 minutes for the droplet case and 60 minutes for the additive manufacturing case. The \ac{3D} rendering process required around 0.1 seconds per time point.

\subsection*{Dataset processing for additive manufacturing}
Standard tomoscopy experiments for additive manufacturing were conducted in a continuous rotation mode while remelting structures that were produced with laser powder bed fusion. 
For each time point, projections within 0-180$\degree$ were captured under the assumption that the sample did not change during this acquisition. 
The raw additive manufacturing dataset, initially sized (400,200,70,528), representing 400 time points and 200 projections over 0-180$\degree$ and each of them with 70x528 pixels, was reshaped to (200,400,70,528) for data extraction over 0-360$\degree$, resulting in half of the time points and double of projections. 
A flat-field correction was applied to reduce background noise. This was followed by phase reconstruction using the method proposed by Paganin et al.~\cite{Paganin2002TIE} to extract the phase information from edge enhancement. 
To further ensure consistency across projections, we leveraged the Radon transform property, which states that the sum of pixel values in each projection remains constant for each time point. 
This approach helps maintain data integrity and improves the accuracy of subsequent reconstructions.

Reconstructions using the conventional tomoscopy scheme with a full 0-180$\degree$ projection range, processed via the Gridrec algorithm~\cite{marone_regridding_2012}, were used as ground truth to evaluate our method. 
The printing process of the additive manufacturing dataset occurred within 40 time points out of a total of 200 time points, so we evaluated only the printing period. 

\section*{Data Availability}
The water droplet collision data that support the findings of this study are available in Figshare with the doi:10.6084/m9.figshare.28533098\cite{figshare2024}. The additive manufacturing data that support the findings of this study are available in the PSI Public Data Repository with the doi: https://doi.org/10.16907/d64d2e8c-b593-47b8-ab90-4ddbd19bedb5\cite{AMfigshare}.
\section*{Code Availability}
The X-Hexplane code is available in the following repository: to be provided when published.

\bibliography{reference}

\section*{Acknowledgments}
Z.H thanks Z. Matej for his support and access to the GPU-computing cluster at MAX IV. 
Z.H, P.V.-P, and M.M. are grateful to F. Marone for the support of the additive manufacturing experiments. 
Z.H, P.V.-P thank R. Klöfkorn for providing the droplet collision simulation dataset.
Z.H, P.V.-P thank C. M. Schlepütz for his review of this article and his insightful comments.
This work has received funding from ERC-2020-STG 3DX-FLASH 948426 and the HorizonEIC-2021-PathfinderOpen-01-01, MHz-TOMOSCOPY 101046448. The Deutsche Forschungsgemeinschaft funded this work through Reinhart-Koselleck project number 408321454, Ba 1170/40
Portions of the text in this manuscript were revised with the assistance of ChatGPT, a large language model developed by OpenAI, to improve clarity and readability. All scientific content was authored and verified by the authors.

\section*{Author Contributions}
Z.H. and P.V.-P. conceived and conceptualized STRT. Z.H., Z. Y., Y.Z., and P.V-P. developed and contributed to the neural network framework and physical formulation of the problem. 
Z.H. and K.J. performed the data analysis. 
M.M. and F. G.-M. provided data and helped to devise experimental scenarios enabled by STRT. 
M.M. also contributed to the detailed analysis of the additive manufacturing results.
P.V.-P. supervised the research.
Z. H. and P. V.-P. wrote the article with input from all the coauthors.

\section*{Competing interests}
The authors declare no competing interests.

\end{document}